\documentclass[prd,twocolumn,aps,amsmath,amssymb,nofootinbib,preprintnumbers]
{revtex4}

\usepackage{graphicx}
\usepackage{dcolumn}
\usepackage{bm}
\usepackage{amsmath}
\usepackage{amsfonts}

\def\ls{\mathrel{\lower4pt\vbox{\lineskip=0pt\baselineskip=0pt
           \hbox{$<$}\hbox{$\sim$}}}}
\def\gs{\mathrel{\lower4pt\vbox{\lineskip=0pt\baselineskip=0pt
           \hbox{$>$}\hbox{$\sim$}}}}
\def\drawbox#1#2{\hrule height#2pt

\hbox{\vrule width#2pt height#1pt \kern#1pt
              \vrule width#2pt}
              \hrule height#2pt}

\def\Asym#1#2{\vcenter{\vbox{\drawbox{#1}{#2}
              \kern-#2pt       
              \drawbox{#1}{#2}}}}

\newcommand{\beq}{\begin{equation}}
\newcommand{\eeq}{\end{equation}}

\begin{document}

\title{Magnetic field production during preheating at the electroweak scale}

\author{Andr\'es D\'\i az-Gil, Juan Garc\'\i a-Bellido, Margarita Garc\'\i a 
P\'erez and Antonio Gonz\'alez-Arroyo}

\affiliation{Instituto de F\'\i sica Te\'orica \ UAM/CSIC,
Universidad Aut\'onoma de Madrid, Cantoblanco, 28049 Madrid, Spain}

\date{December 27, 2007}

\begin{abstract}
We study the generation of magnetic fields during preheating within an
scenario of hybrid inflation at the electroweak (EW) scale. We find
that the non-perturbative and strongly out-of-equilibrium process of
magnetic field production occurs along the lines predicted by
Vachaspati many years ago. The system starts in the false vacuum at
the end of inflation, and very quickly the initial quantum
fluctuations of the Higgs field get amplified via long wavelength
spinodal instabilities.  The subsequent nucleation of the random
Gaussian Higgs field bubbles (lumps) leads to EW symmetry breaking,
and to the creation of $Z$-strings, which soon decay, along with
longwave magnetic flux tubes with nontrivial helicity. The intensity
and scales in these helical magnetic fields are consistent with their
later development into the microgauss fields observed in galaxies and
clusters of galaxies.
\end{abstract}
\preprint{IFT-UAM/CSIC-07-65}
\maketitle


The origin of primordial magnetic fields (PMF) is one of the remaining
puzzles in cosmology~\cite{Reviews}.  Magnetic fields play an
important role in the evolution of the primordial plasma in the early
universe, and in the propagation of high energy cosmic rays in our
galaxy and beyond.  They may influence galaxy and large scale
structure formation, and may generate a stochastic background of
gravitational waves.

Magnetic fields have been measured on the scales of galaxies and
clusters of galaxies with a magnitude of order
microgauss~\cite{Vallee1997}.  There is even some evidence of their
presence on scales of superclusters, and associated with quasars at
redshifts $z\sim 2$. The main difficulty in understanding their origin
is not in their magnitude but in its correlation scale. The $\mu$G
order of magnitude of galactic MF could be explained by an
amplification via a dynamo mechanism initiated by a tiny seed,
$B\sim10^{-30}$ G, in the context of gravitational collapse in a flat
$\Lambda$CDM model. Their magnitude on clusters scales is much more
difficult to explain via a dynamo mechanism~\cite{Reviews}. In any
case, an initial seed will be required. There are both astrophysical
and cosmological theoretical models trying to account for the origin
of the primordial seed: Biermann battery in intergalactic shocks,
stellar magnetic winds (like in our Sun), supernova explosions,
galactic outflows in the inter-galactic medium (IGM), quasar outflows
of magnetized plasma into the intra-cluster medium
(ICM)~\cite{Reviews}; as well as from early universe phase
transitions~\cite{PhaseTrans}, from magnetic helicity at the EW
transition, together with the baryon asymmetry of the universe
(BAU)~\cite{Vachaspati}, via hypercharge and hypermagnetic field
generation before the EW transition~\cite{Shaposhnikov}, from second
order cosmological perturbations from inflation~\cite{CosmicPert},
from reheating after inflation~\cite{Reheating}, or produced during
inflation~\cite{Inflation}, etc.

The ubiquity of microgauss MF on all scales suggests a cosmic origin.
Could they be generated in the early universe and then redshifted
until today? Magnetic fields could have appeared as the electroweak
force broke into  weak interactions plus electromagnetism, at a
scale $T_{\rm EW} \sim 100$ GeV, when the universe had an energy
density $\rho_{\rm EW} \sim 10^8$ GeV$^4$. The universe was (or
became) radiation dominated at that time, and the energy density carried  by magnetic
fields could correspond to a fraction $f$ of the total. If we redshift this energy
density until today ($T_0 = 2.725$ K) we get $\rho_{\rm B}(t_0) =
(T_0/T_{\rm EW})^4\ f\ \rho_{\rm EW} \sim 3\times10^{-53}\ f\ {\rm
GeV}^4 = 0.4\ f\,{\rm eV/cm}^3 = (5\ f^{1/2}\, \mu{\rm G})^2/(8\pi)$,
where we have used 1 G $= 1.95\times10^{-20}$ GeV$^2$.  This would be
enough to explain the cluster and supercluster values, and would
perhaps require a mild dynamo mechanism to amplify it to galactic
values (if the fraction $f\ll 1$). 

However, a priori, it is not so clear how one can obtain the large
correlation length of magnetic fields observed at galactic and cluster
scales. Any physical mechanism that creates magnetic fields must
necessarily be causal, but at high temperatures in the early universe
there is also a natural coherence scale given by the physical
horizon. At the EW scale the particle horizon is $10^{-10}$
lightseconds ($\sim$~3~cm), which today corresponds to a comoving
scale of $\sim$~1~AU, clearly insufficient when compared even with the
irregular (turbulent) component of the galactic magnetic field ($L
\sim 100$ parsecs), not to mention the regular (uniform) component,
which has correlations $L \sim 10$ kpc. It thus seems apparently
impossible to explain with our mechanism the coherent magnetic fields
observed on galaxy clusters and supercluster scales (of order 10 Mpc)
with intensities of order $\mu$G to $n$G.  Nevertheless, if we assume
that the plasma after the EW transition is sufficiently turbulent to
maintain magnetic fields of the largest possible coherence scales via
inverse cascade, then we could reach cosmological scales today.  The
largest coherence scale is the physical horizon.  If a strong inverse
cascade is active, then the coherence length of the magnetic fields
could grow as fast as the horizon, i.e.  like the scale factor {\em
squared} during the radiation dominated era.  This optimal situation
is only attainable in the presence of a plasma, and thus is bound to
stop acting at photon decoupling, when the universe becomes
neutral. Since then, the correlation length can only grow with the
expansion of the universe, as the scale factor. Taking these facts
into account, and using the adiabatic expansion relation, we can
compute the coherence scale of the field in terms of that at the
electroweak scale,
$\ \xi_0 = \xi_{\rm
EW}\,\left({a_{\rm dec}\over a_{\rm EW}}\right)^2 {a_0\over a_{\rm
dec}} = 3\,{\rm cm} \left({T_{\rm EW}\over T_{\rm eq}}\right)^2{T_{\rm
eq}\over T_0} \sim 6\times 10^{25}\,{\rm cm} = 20\ {\rm Mpc}$, where
we have made the approximation that matter-radiation equality and
photon decoupling occurred more or less simultaneously (a careful
computation gives only a minor correction).  The surprising thing is
that this simple calculation gives precisely the order of magnitude
for the largest correlation length of cosmic MF ever observed
(i.e. cluster scales). If the $\mu$G magnitude of the primordial MF
seed, arising from simple scaling down the energy density since the EW
transition, seemed a curious coincidence, the fact that an inverse
cascade could also be responsible for the observed correlation length
becomes a suspicious coincidence.  These observations have triggered a
large number of investigations devoted to the analysis of
magnetogenesis during the electroweak phase
transition~\cite{PhaseTrans,Vachaspati,Shaposhnikov}.

In this Letter we address the question of whether the required
conditions --a significant fraction of energy density in magnetic
fields with a large helical component and sustained periods of inverse
cascade -- can be obtained within a simple and concrete scenario of
low-scale hybrid inflation~\cite{CEWB}. We analyze whether large scale
magnetic fields are generated during the preheating and reheating
periods which lead to EW symmetry breaking. Previous attempts to
analyze magnetogenesis during reheating as a consequence of spinodal
instability can be found in Ref.~\cite{Reheating}. Our approach differs
significantly and is based on the use of the classical approximation:
integration of the non-linear equations of motion with stochastic
initial conditions determined by matching the quantum expectation
values at the end of inflation. The numerical implementation is based
upon discretization of our system onto a space-time lattice. Our
results indicate that a mechanism resembling that introduced by
Vachaspati~\cite{Vachaspati} could have taken place in the early
universe. The mechanism relies on the fact that an inhomogeneous Higgs
background may act as a source for magnetic fields. The effect would
come from the existence of non-trivial gradients of the Higgs field
phase which would generate $W$-currents and magnetic fields during a
cold EW transition~\cite{CEWB}.  These gradients are usually
associated with the appearance of unstable $Z$-strings, together with
the helical magnetic field necessary for growth into large scales via
MHD turbulence~\cite{Vachaspati}.  As we will see, our results show
that during the initial stage of electroweak symmetry breaking (EWSB),
our setup is indeed characterized by the appearance of $Z$-strings
and magnetic strings with non-trivial helicity.

\begin{figure}[t]
\begin{center}
\includegraphics[width=5.5cm,angle=0]{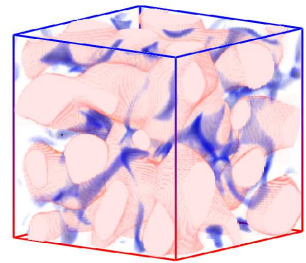}
\end{center}
\vspace*{-5mm}
\caption{The location of the Higgs lumps (light/red) and the magnetic field
flux tubes (dark/blue) at $mt=15$.}
\label{fig1}
\vspace*{-3mm}
\end{figure}

We consider a specific model of Hybrid inflation with the field
content of the gauge plus scalar sector of the Standard Model coupled
to a singlet scalar field, the inflaton~\cite{CEWB}. The gauge sector
includes both the SU(2) and the hypercharge U(1) fields with
corresponding gauge couplings $g_{\rm W}$ and $g_{\rm Y}$.  The scalar
sector contains the 4-real component Higgs field $\phi_\mu(x)$, and a
singlet inflaton field $\chi(x)$ coupled to the Higgs via the term
$g^2\,\phi^2_\mu\,\chi^2$.  Our initial condition assumes that at the
end of inflation both the Higgs and the gauge fields are in the de
Sitter vacuum. In fact, since for EW-scale inflation the rate of
expansion is negligible compared to any other scale, $H\approx
10^{-5}$ eV $\ll v=246$ GeV, the de Sitter vacuum state is
indistinguishable from the Minkowski vacuum for the range of momenta
we are considering.  The inflaton field is given by the homogeneous
mode alone, which is decreasing with time.  This time dependence
translates, through the Higgs-inflaton coupling, into a time
dependence of the Higgs mass-squared, which changes from positive to
negative inducing an exponential growth of the Higgs low momentum
modes in a process known as ``tachyonic preheating''~\cite{GBKLT}.
The remaining non-tachyonic fields are populated through their
interaction to the Higgs. Fermionic fields can be safely ignored from
the analysis since the perturbative decay of the Higgs into quarks and
leptons occurs within a time scale $mt\sim10^3-10^4$, many times
bigger than the scales we are considering in our simulations.

There have been several numerical studies of this
model~\cite{GBKLT,jmt}, in an attempt to explain the BAU in this
context~\cite{CEWB,jmt2,Smit}, as well as the production of a
gravitational wave background~\cite{GWpreh}.  In this paper, the
Hypercharge gauge field is included for the first time(preliminary
results were discussed in Ref.\cite{ajmt}).  This will allow us to
study the generation of U(1) electromagnetic fields during preheating.
In order to analyze the results we first have to provide a definition
of the U(1)$_{\rm em}$ content of the SU(2)$\times$U(1) fields in the
Lagrangian. We will be using a gauge invariant definition of $Z$
vector potential on the lattice as
\begin{equation}
\hat Z_\mu (n) = {{\rm Tr}\Big[ - i \hat n 
\hat D_\mu (\hat \Phi(n)) \hat \Phi^\dagger (n) \Big]
\over 2|\hat \phi(n)| |\hat \phi(n+\mu)|}  
\,, \label{zeta}
\end{equation}
where $\hat\phi_\mu(n)$ are the  lattice Higgs field components, 
with modulus $|\hat \phi(n)|$. In matrix notation,
$\hat \Phi(n)= \phi_0\mathbf{I} + i \phi_a \tau_a$. We have introduced
the adjoint unit vector $\hat n = n_a \tau_a$, with components $n_a(n)
= \varphi^\dagger (n) \tau_a \varphi(n) / |\varphi(n)|^2$, with
$\varphi(n) = \Phi(n) (1,0)^{\rm T}$ the Higgs doublet, and $\hat D_\mu$ is
the SU(2)$\times$U(1) lattice covariant derivative.  Note that our definition
of the $Z$ vector potential corresponds to the standard one in the
unitary gauge.  We can then compute the field strength of the U(1)$_{\rm
  em}$ field in terms of the $Z$ and Hypercharge field strengths:
\begin{equation}\label{fem}
\hat F^{\gamma}_{\mu \nu}(n)  = \sin^2(\theta_W)
\hat F^{\rm Z}_{\mu \nu}(n) - \hat  F^{\rm Y}_{\mu \nu}(n) \,.
\end{equation}
This procedure preserves the U(1)$_{\rm em}$  gauge-invariance 
on the lattice.

\begin{figure}[t]
\begin{center}
\includegraphics[width=4.2cm,angle=0]{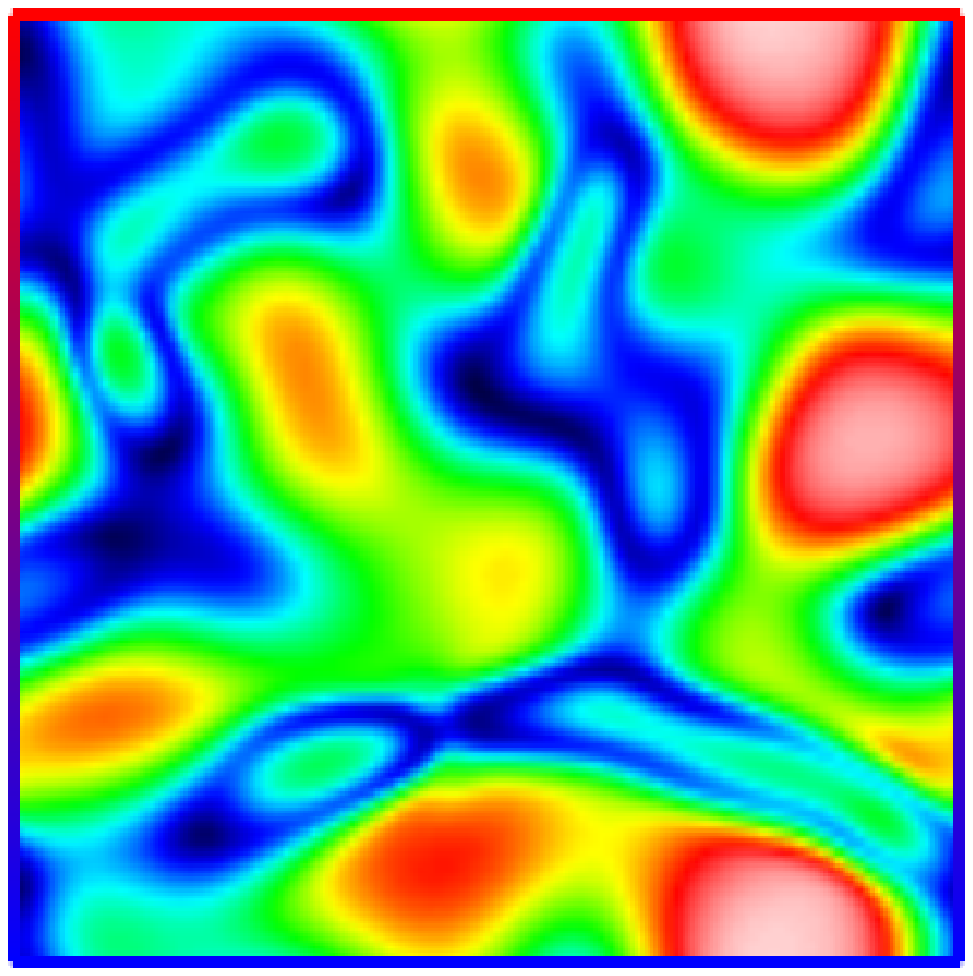}
\includegraphics[width=4.2cm,angle=0]{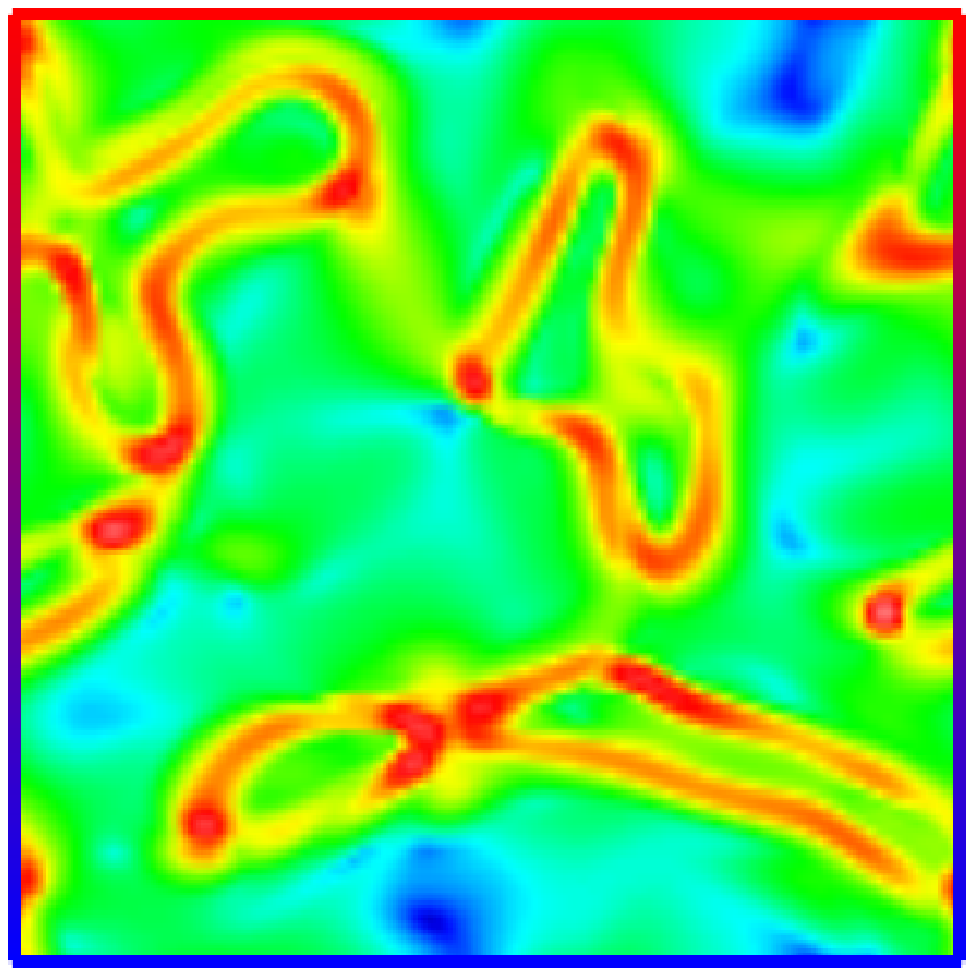}
\hspace*{-2mm}
\includegraphics[width=3.7cm,angle=0]{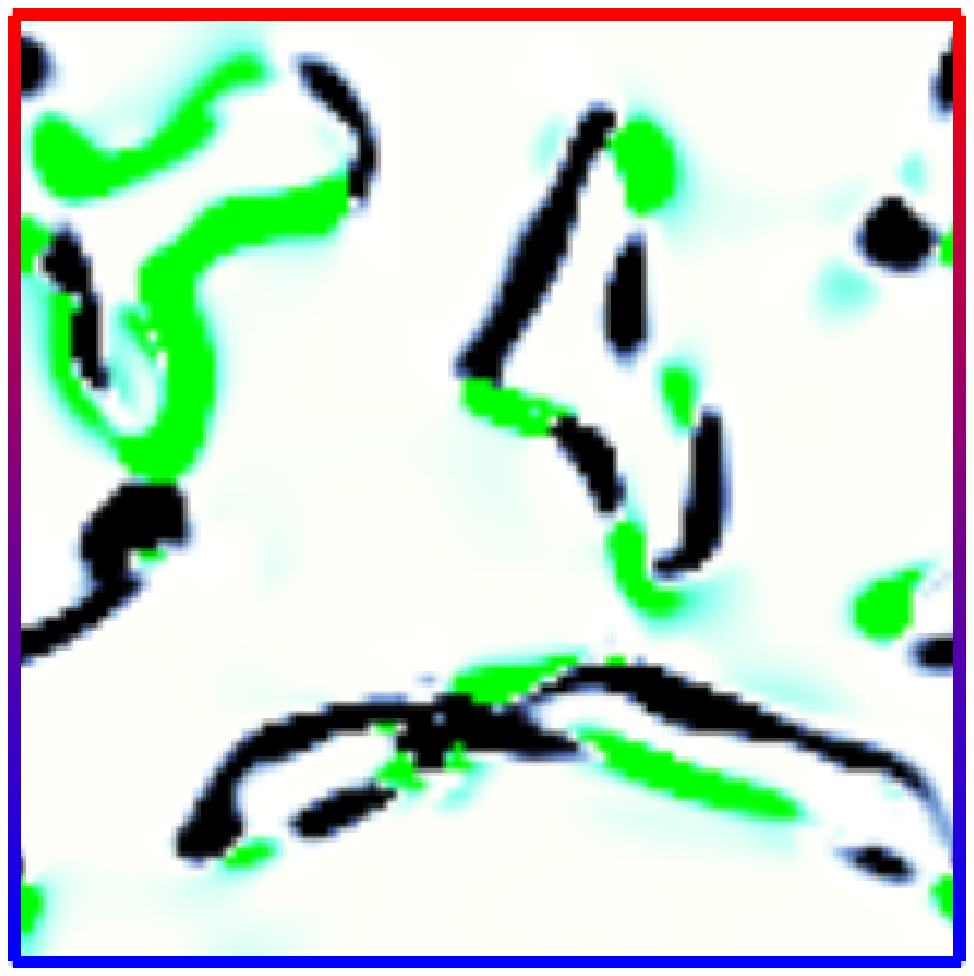}
\hspace*{4.5mm}
\includegraphics[width=3.6cm,angle=0]{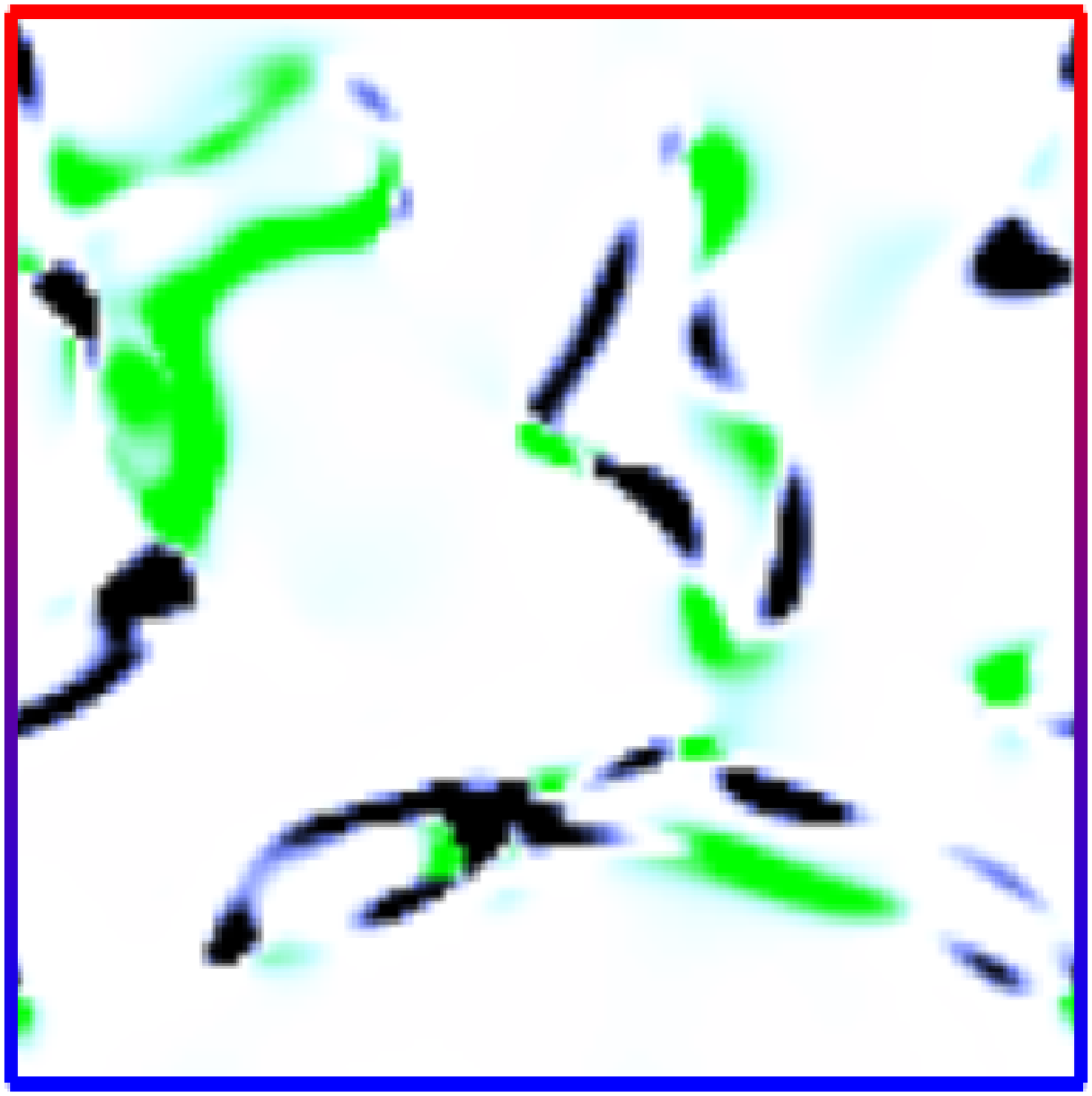}
\end{center}
\vspace*{-5mm}
\caption{2D contour plots of the Higgs norm (top left), magnetic field
flux tubes (top right), $Z$-string helicity (bottom left) and 
magnetic field helicity (bottom right) at $mt=15$. 
Note the precise spatial correlation.}
\label{fig2}
\vspace*{-3mm}
\end{figure}

As mentioned previously, there are two essential ingredients in
Vachaspati's proposal for magnetogenesis: the existence of
inhomogeneities in the Higgs field and the generation of a non-trivial
magnetic field helicity.  Both of them are present in the first stages
of evolution right after inflation ends. This might seem surprising
since initially both hypercharge and SU(2) magnetic fields are
zero. The way it comes about is by noticing that local variations of
the orientation of the Higgs field induce via Eq.~(\ref{fem}) a
non-vanishing contribution to the electromagnetic field.  The initial
conditions for the classical evolution are given by a Higgs field
as a Gaussian random field.  This initial distribution carries
essential information about how the process of EW symmetry breaking
takes place. It is, for instance, the seed for the growth of bubble
shells which arise from lumps in the initial distribution of the Higgs
field norm. We have observed that it also determines the way in which
magnetic fields are seeded.  At later times, after EWSB has occured,
the correlation between the Higgs field norm and magnetic fields is
manifest.  In Fig.~\ref{fig1} we show a snapshot of both quantities
at a time ($mt=15$) after EWSB has occurred and bubble shells (in red)
fill almost all the volume of the box.  Magnetic fields (in blue) are
localized in string-like structures concentrated in the region between
bubbles.  It is there where the Higgs field remains closer to the
false vacuum for a longer period of time giving rise to larger
gradients in the Higgs field phases.  This linkage between magnetic
strings and Higgs field minima is even more evident in the two
dimensional contour plots presented in Fig.~\ref{fig2}. As mentioned
previously part of this structure was already present in our initial
condition ($mt=5$), in the absence of SU(2) and hypercharge magnetic
fields, but it has become far more complex at this stage.

It is interesting to point out that previous analysis of the cold EW
transition, neglecting the effect of hypercharge, have revealed that
regions between bubbles are also sources of Chern-Simons number
through the appearance of sphaleron-like configurations attached to
the location of zeroes in the Higgs field norm~\cite{jmt2,Smit}. Since
for non-zero Weinberg angle sphalerons are like magnetic dipoles, it
is tempting to correlate the observed U(1) magnetic strings to the
alignment of sphaleron dipoles. A detailed investigation of this issue
is beyond the scope of this paper. In any case, the connection between
Chern-Simons number and magnetic structures is more explicit when
analyzing the generation of magnetic helicity, which can be regarded
as the Abelian counterpart of the Chern-Simons number.  During these
first stages of evolution the magnetic and $Z$-boson helicities are
indeed tracking the magnetic strings (a clear example is shown in
Fig.~\ref{fig2}, comparing top and bottom boxes).  The proportionality
of the $Z$-boson and photon helicities in the initial condition is a
direct consequence of our definition of the fields. Nonetheless this
correlation is still preserved once gauge fields and non-linearities
have started to play a role.


\begin{figure}[t]
\vspace*{-19mm}

\begin{center}
\includegraphics[height=8cm,angle=-90]{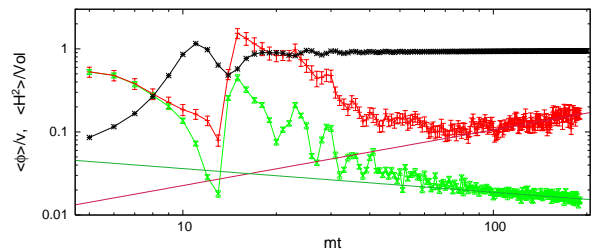}
\end{center}
\vspace*{-3mm}
\caption{Time evolution of the Higgs v.e.v. (black) and the square of
  the helicity in $Z$ (green) and U(1) (red) magnetic fields
  normalized to the physical volume ${\cal V}$.  The $Z$-helicity 
  has been rescaled by $\tan^2\theta_W$.  For $m_{\rm h} = 2m_{\rm
  w}$, the corresponding slopes are: Photon field: $0.68(7)$ and $Z$
  field: $-0.27(4)$.}
 \label{fig3}
 \vspace*{-3mm}
 \end{figure}

The previous figures correspond to a snapshot at $mt=15$, when the
string structure comes out to be stronger. It is important to
ascertain whether the later stages of evolution preserve relic
magnetic fields with non trivial magnetic helicity.  This quantity is
defined as:
\begin{equation}
H =   \int d^3 x  \vec A \cdot \vec B \equiv {1\over {\cal V} } \sum_k  {-i \vec
k \over |\vec k|^2} \cdot (\vec B(\vec k) \times \vec B^*(\vec k)) \,,
\label{helicity}
\end{equation}
where ${\cal V}$ is the volume of space. Being a pseudoscalar and
since we have not included CP violation, the expectation value of the
helicity is zero. Nevertheless, to have an idea of the efficiency of
this mechanism in generating non-trivial magnetic helicity we have
computed its dispersion. Its time evolution is presented in
Fig.\ref{fig3} where we also present the time dependence of the
helicity dispersion in the $Z$-boson magnetic field. The latter has been
rescaled by $\tan^2\theta_W$ to make it agree with the electromagnetic
helicity for the initial configuration at $mt=5$.  While initially
both helicities coincide, time evolution damps the $Z$-helicity while
it enhances the Abelian magnetic one. This agrees with Vachaspati's
picture, that predicted an initial correlation among both quantities,
and subsequent decay of $Z$-strings, leaving remnant magnetic fields
with non-trivial helicity. From our analysis we determine that the
magnetic helicity dispersion is actually growing in time with a
power-law behavior, $\sim t^{0.68(7)}$. A detailed study of the late
time behavior of the system will be reported in Ref.~\cite{ajmt2}. We
will discuss there the nature of the plasma during the first stages of
reheating which shows evidence of electric charge separation. We will
also present some evidence of the presence of inverse cascade in the
magnetic power spectrum and the helicity dispersion.

In summary, in this paper we studied the preheating and early
reheating periods of a model of hybrid inflation at the EW scale
within the classical approximation. This is the first study of this
sort including SU(2)$\times$U(1). We have focused upon the generation
of magnetic fields in this context and provided evidence that the
mechanism proposed in 1991 by Vachaspati~\cite{Vachaspati} is at
work. Spatial gradients of the Higgs field, seeded by those already
present in the random Gaussian initial condition, act as sources of
the magnetic field. The latter has a non-trivial helicity dispersion,
correlated initially with the presence of $Z$-strings, that is growing
with time.

{\it Acknowledgements -} It is a pleasure to thank A. Ach\'ucarro,
M. Hindmarsh, M. Salle, M. Shaposhnikov, J. Smit, J. Urrestilla and
T. Vachaspati for generous discussions.  We acknowledge financial
support from the Madrid Regional Government (CAM) under the program
HEPHACOS P-ESP-00346, and the Spanish M.E.C. under contracts
FPA2006-05807, FPA2006-05485 and FPA2006-05423. We also acknowledge
the use of the MareNostrum supercomputer and the IFT computation
cluster.


\end{document}